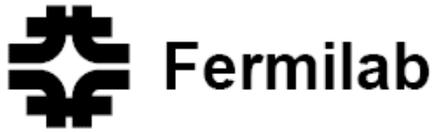

# RADIONUCLIDES IN THE COOLING WATER SYSTEMS FOR THE NUMI BEAMLINE AND THE ANTIPROTON PRODUCTION TARGET STATION AT FERMILAB[†]

Hiroshi Matsumura [a*], Shun Sekimoto [b], Hiroshi Yashima [b], Akihiro Toyoda [a], Yoshimi Kasugai [c], Norihiro Matsuda [c], Koji Oishi [d], Kotaro Bessho [a], Yukio Sakamoto [c], Hiroshi Nakashima [c], David Boehnlein [e], Gary Lauten [e], Anthony Leveling [e], Nikolai Mokhov [e], Kamran Vaziri [e]

[a] *High Energy Accelerator Research Organization (KEK), 1-1 Oho, Tsukuba, Ibaraki 305-0801 Japan;* [b] *Kyoto University Research Reactor Institute, 2 Asashiro-Nishi, Kumatori-cho, Sennan-gun, Osaka 590-0494, Japan;* [c] *Japan Atomic Energy Agency, 2-4 Shirane Shirakata, Tokai-mura, Naka-gun, Ibaraki, 319-1195, Japan;* [d] *Shimizu Corporation, 3-4-17 Etchujima, Koto-ku, Tokyo, 135-8530, Japan;* [e] *Fermi National Accelerator Laboratory (Fermilab), Batavia, IL 60510-5011, USA*

## Abstract

At the 120-GeV proton accelerator facilities of Fermilab, USA, water samples were collected from the cooling water systems for the target, magnetic horn1, magnetic horn2, decay pipe, and hadron absorber at the NuMI beamline as well as from the cooling water systems for the collection lens, pulse magnet and collimator, and beam absorber at the antiproton production target station, just after the shutdown of the accelerators for a maintenance period. Specific activities of γ-emitting radionuclides and $^3$H in these samples were determined using high-purity germanium detectors and a liquid scintillation counter. The cooling water contained various radionuclides depending on both major and minor materials in contact with the water. The activity of the radionuclides depended on the presence of a deionizer. Specific activities of $^3$H were used to estimate the residual rates of $^7$Be. The estimated residual rates of $^7$Be in the cooling water were approximately 5% for systems without deionizers and less than 0.1% for systems with deionizers, although the deionizers function to remove $^7$Be from the cooling water.

# Radionuclides in the Cooling Water Systems for the NuMI Beamline and the Antiproton Production Target Station at Fermilab

Hiroshi Matsumura [a*], Shun Sekimoto [b], Hiroshi Yashima [b], Akihiro Toyoda [a], Yoshimi Kasugai [c], Norihiro Matsuda [c], Koji Oishi [d], Kotaro Bessho [a], Yukio Sakamoto [c], Hiroshi Nakashima [c], David Boehnlein [e], Gary Lauten [e], Anthony Leveling [e], Nikolai Mokhov [e], Kamran Vaziri [e]

[a] High Energy Accelerator Research Organization (KEK), 1-1 Oho, Tsukuba, Ibaraki 305-0801 Japan; [b] Kyoto University Research Reactor Institute, 2 Asashiro-Nishi, Kumatori-cho, Sennan-gun, Osaka 590-0494, Japan; [c] Japan Atomic Energy Agency, 2-4 Shirane Shirakata, Tokai-mura, Naka-gun, Ibaraki, 319-1195, Japan; [d] Shimizu Corporation, 3-4-17 Etchujima, Koto-ku, Tokyo, 135-8530, Japan; [e] Fermi National Accelerator Laboratory (Fermilab), Batavia, IL 60510-5011, USA

At the 120-GeV proton accelerator facilities of Fermilab, USA, water samples were collected from the cooling water systems for the target, magnetic horn1, magnetic horn2, decay pipe, and hadron absorber at the NuMI beamline as well as from the cooling water systems for the collection lens, pulse magnet and collimator, and beam absorber at the antiproton production target station, just after the shutdown of the accelerators for a maintenance period. Specific activities of γ-emitting radionuclides and $^3$H in these samples were determined using high-purity germanium detectors and a liquid scintillation counter. The cooling water contained various radionuclides depending on both major and minor materials in contact with the water. The activity of the radionuclides depended on the presence of a deionizer. Specific activities of $^3$H were used to estimate the residual rates of $^7$Be. The estimated residual rates of $^7$Be in the cooling water were approximately 5% for systems without deionizers and less than 0.1% for systems with deionizers, although the deionizers function to remove $^7$Be from the cooling water.

## 1. INTRODUCTION

High-energy accelerator components that are cooled by circulating water are exposed to both primary beam protons and secondary particles. As a result, some of the radionuclides are produced directly in the cooling water, while some are produced in the accelerator components and are transferred to the cooling water by physical and/or chemical processes [1, 2]. At J-PARC in Japan [3], cooling water systems for high-energy accelerator components are scheduled to be drained and refilled at regular intervals. Radionuclide concentrations in the cooling water are required to be below the regulatory limits to allow these cooling water systems to be drained. Therefore, to predict the activity levels and behavior of the radionuclides in the cooling water, radioactive water in these systems had to be characterized before regular machine operation.

Fermilab in the USA has high-energy high-intensity accelerators. In order to determine the activity levels and behavior of the radionuclides in operating cooling water systems, in this study, specific activities of the radionuclides were determined for cooling water samples obtained from eight systems for the 120-GeV proton beamlines at Fermilab. The water from five cooling water systems in the NuMI beamline (NuMI) and from three ones in the antiproton production target station (Pbar) shown in **Figure 1** was investigated. Furthermore, residual rates of $^7$Be in the cooling water were estimated from the measured $^3$H and $^7$Be activities.

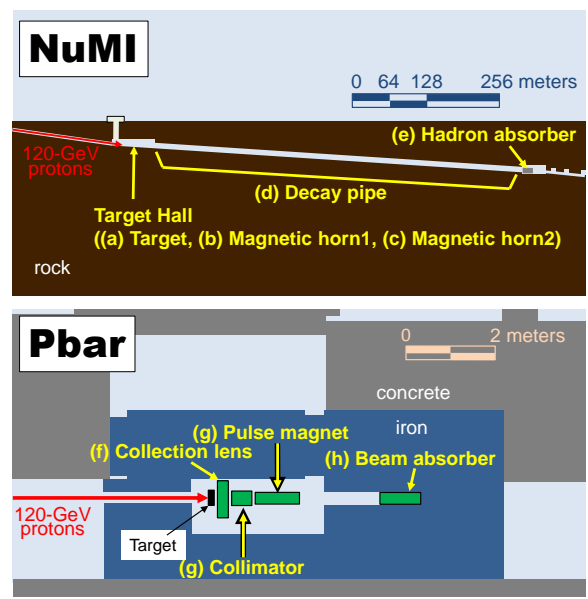

Figure 1. Accelerator components with cooling-water at NuMI (top) and Pbar (bottom) in Fermilab.

*Corresponding author. Email: hiroshi.matsumura@kek.jp

Table 1. Information on the cooling-water systems.

| System name | Water volume | Filter | Deionizer | Water contact materials |
|---|---|---|---|---|
| (a) NuMI, Target | 133 L | Yes | Yes | Different grades of stainless steel and brazing material, 305L stainless steel pipe (outside) |
| (b) NuMI, Magnetic horn1 | 190 L | Yes | Yes | Nickel coating and anodization should keep water away from 6061 Al alloy horn, 305L stainless steel pipe (outside) |
| (c) NuMI, Magnetic horn2 | 228 L | Yes | Yes | Nickel coating and anodization should keep water away from 6061 Al alloy horn, 305L stainless steel pipe (outside) |
| (d) NuMI, Decay pipe | 2755 L | Yes | Yes | Soft copper (on decay pipe), Type K hard copper pipe (outside) |
| (e) NuMI, Hadron absorber | 513 L | Yes | Yes | Al tube (absorber module) |
| (f) Pbar, Collection lens | 60 L | Yes | Yes | Cu pipe, stenless steel pipe (outside), Ti alley (lens surface) |
| (g) Pbar, Pulse magnet and collimator | 60 L | Yes | No | Cu pipe |
| (h) Pbar, Beam absorber | 60 L | Yes | No | Cu pipe |

## 2. EXPERIMENTAL PROCEDURES

### 2.1. Cooling water systems

In NuMI, we analyzed the cooling water for (a) the target, (b) magnetic horn1, (c) magnetic horn2, (d) decay pipe, and (e) hadron absorber. In Pbar, we investigated the cooling water for (f) the collection lens, (g) pulse magnet and collimator, and (h) beam absorber. Those are illustrated in Figure 1. In this analysis, 120-GeV proton beams were transported to NuMI and Pbar.

Important information on the cooling water systems is listed in **Table 1**. Total water volumes are different among these systems, ranging from 60 to 27,552 L. Each system has a filter unit. Further, the cooling water systems for (a), (b), (c), (d), (e), and (f) are equipped with a deionizer that contains a mixture of cation-exchange resin and anion-exchange resin. The cooling water systems for (g) and (h) are without deionizers. Water-contacting materials are also different among the systems. Typical frequencies of water exchange are twice per year for (b), (c), (g), and (h), once every 1–2 years for (a) and (f), and never for (d). 1/3 of the volume of water was exchanged in fall 2007 in the system for (e).

### 2.2. Cooling water sampling

In February 2010, December 2010, and September 2011, 500 mL of cooling water from the eight cooling water systems listed in Table 1 were collected in polyethylene bottles just after the accelerators were shut down for maintenance. Cooling water samples for (a), (b), (c), and (d) was not collected in February 2010.

### 2.3. Gamma-ray spectrometry

From the sample bottle of the original cooling water, 100 mL or 50 mL of water was transferred to a 100-mL polypropylene bottle for measurement. The sample volumes were determined by weight. In order to acidify the water, 0.100 mL of 6 M $HNO_3$ was added to the polypropylene bottle. The radioactivity in the water at the end of the machine operation was determined by γ-ray spectrometry.

In order to measure γ-rays from anion species of radionuclides, 100 mL of the original cooling water collected in September 2011 was passed through a cation-exchange resin column (Dowex 50WX8, 100–200 mesh, $H^+$ form, 1 mL volume). The passed water was transferred to a 100-mL polypropylene bottle. The radioactivity in the water at the end of machine operation was determined by γ-ray spectrometry.

The γ-ray spectrometry was performed with high-purity germanium (HPGe) detectors. One of the HPGe detectors was previously calibrated by Canberra [4]. The efficiencies of the previously calibrated detector were determined using Canberra's LabSOCS software [5]. The detection efficiencies of the other HPGe detectors were determined from the ratios of the peak counting rate of the calibrated HPGe detector to that of the uncalibrated HPGe detectors. Nuclear data in the literature [6] were used for data analysis.

### 2.4. Tritium counting

A fraction of the original cooling water samples was diluted to 1/100 concentration and was used to measure $^3H$ activity. Then 2.5 mL of the diluted water was mixed with 10 mL of a scintillation cocktail. Radioactivity of $^3H$ was determined by a liquid scintillation counter.

Table 2. Detected radionuclides in the cooling water and its specific activity. The details are described in text.

| System name | Detected radionuclides and specific activities (Bq/cm$^3$) | | | | | |
|---|---|---|---|---|---|---|
| (a) NuMI, Target | $^3$H 2.69k | $^7$Be 3.8 | $^{24}$Na 0.03 | $^{52}$Mn 0.005 | $^{54}$Mn 0.005 | $^{60}$Co 0.003 |
| (b) NuMI, Magnetic horn1 | $^3$H 11.3k | $^{39}$Cl 2.8k | $^{24}$Na 0.026k | $^{38}$S 0.017k | $^{38}$S 0.0118k | $^{28}$Mg 0.9 | $^{22}$Na 0.04 |
| (c) NuMI, Magnetic horn2 | $^3$H 1.97k | $^{39}$Cl 1.07k | $^{38}$S 3.1 | $^{24}$Na 1.9 | $^7$Be 0.33 | $^{28}$Mg 0.09 | $^{22}$Na 0.006 |
| (d) NuMI, Decay pipe | $^3$H 3.96k | $^7$Be 1.6 | $^{52}$Mn 0.010 | $^{58}$Co 0.008 | $^{54}$Mn 0.004 | | |
| (e) NuMI, Hadron absorber | $^3$H 7.40k | $^7$Be 3.5 | $^{24}$Na 0.21 | $^{82}$Br 0.063 | | | |
| (f) Pbar, Collection lens | $^3$H 9.18k | $^7$Be 1.8 | | | | | |
| (g) Pbar, Pulse magnet and collimator | $^3$H 25.2k | $^7$Be 6.1k | $^{58}$Co 1.18k | $^{57}$Co 0.93k | $^{54}$Mn 0.48k | $^{60}$Co 0.35k | $^{56}$Co 0.21k |
| | $^{65}$Zn 0.096k | $^{52}$Mn 0.094k | $^{24}$Na 0.090k | $^{43}$K 0.09k | $^{22}$Na 0.028k | $^{182m}$Re 2.1 | $^{181}$Re 2.0 |
| (h) Pbar, Beam absorber | $^3$H 18.5k | $^7$Be 6.5k | $^{57}$Co 0.60k | $^{58}$Co 0.53k | $^{54}$Mn 0.41k | $^{60}$Co 0.34k | $^{56}$Co 0.094k |
| | $^{43}$K 0.059k | $^{24}$Na 0.054k | $^{52}$Mn 0.042k | $^{22}$Na 0.033k | $^{65}$Zn 0.025k | $^{182m}$Re 1.4 | $^{181}$Re 1.4 |

## 3. RESULTS AND DISCUSSION

### 3.1. Detected radionuclides

The detected radionuclides in the cooling water and its specific activities are listed in **Table 2**. The radionuclides are arranged in order of specific activity. Here, the non-underlined radionuclide term and values were measured in water collected in December 2010. The underlined radionuclide term and values were measured in water (collected in September 2011) passing through a cation-exchange resin column. Many radionuclides were detected in the cooling water. As a common feature, activities of $^3$H and $^7$Be were significantly higher than those of others although the half-lives of $^3$H and $^7$Be are relatively long. This result indicated that most of the $^3$H and $^7$Be was produced directly in the cooling water by nuclear spallation of oxygen in water molecules.

The existence of other radionuclides depended on the cooling water system. In the sample for (a), medium- mass radionuclides were observed. Those might have been emitted and/or released from the stainless steel. In the samples for (b) and (c), lighter-mass radionuclides were observed. Those might have been from the Al alloy. In the sample for (d), spallation radionuclides from Cu were observed. In the samples for (e), $^{24}$Na and $^{82}$Br were observed. Although the source of $^{24}$Na must have been Al, $^{82}$Br is of unknown origin. In the samples for (f), no other radionuclide except for $^3$H and $^7$Be was observed. In the samples for (g) and (h), many spallation radionuclides from Cu were observed at high specific activity. The high specific activity might be due to the lack of a deionizer system. Re isotopes were observed in the samples passed through the cation-exchange column. Those are also of unknown origin.

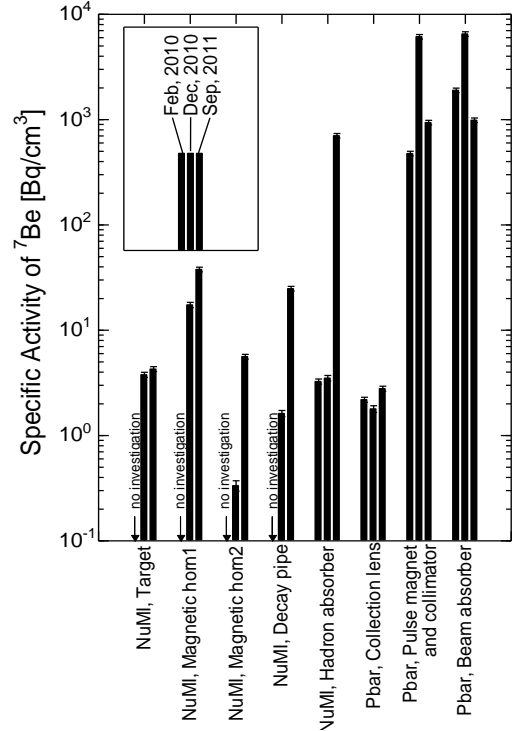

Figure 2. Comparison of the $^7$Be-specific activities in the cooling water collected from NuMI and Pbar in February 2010, December 2010, and September 2011.

The observed variety of the radionuclides in the cooling water indicated that the behavior of radionuclides in the cooling water complicatedly depends on both major and minor materials in contact with the water. Among the γ-emitting radionuclides with half-lives longer than a few days, $^7$Be has priority over others in radioactivity control because of its extremely high specific activity.

### 3.2. Specific activity of $^7$Be in the cooling water

**Figure 2** shows comparison of the $^7$Be-specific activity in the cooling water samples collected from NuMI and Pbar in February 2010, December 2010, and September 2011. The specific activity is independent of sampling date except for that from the NuMI hadron absorber on September 2011. The cooling water systems could be separated into a lower (the left 6 systems in Figure 2) and a higher (the right 2 systems in Figure 2) groups for $^7$Be-specific activity. The lower group was equipped with a deionizer, while the higher group was not. It was assumed that the deionizers reduced the activity of $^7$Be in the water. By monitoring water conductivity, it was found that the deionizer in the NuMI hadron absorber was saturated with ions from April 2011. Because the deionizer in the NuMI hadron absorber could not adsorb any more $^7$Be, the specific activity of $^7$Be might have increased significantly from December 2010 to September 2011.

From these results, we assumed that $^7$Be was removed from the water by the deionizer. However, total water volume and particle flux are different among the systems. Therefore, in the next section, residual rates of $^7$Be in the cooling water were estimated in order to clarify the deionizer function.

### 3.3. Residual rates of $^7$Be in the cooling water

Most of the $^3$H and $^7$Be activities were produced directly in the cooling water as described in Section 3.1. With $^3$H loss by water decomposition and leakage being ignored, the residual rates of $^7$Be in the water were estimated using Eq. (1) because $^3$H is not removed from the cooling water by the deionizer:

$$\text{Residual rate of }^7\text{Be} = \left[\frac{A_{\text{Be-7}}}{A_{\text{H-3}}}\right]_m \cdot \left[\frac{\sigma_{\text{H-3}}}{\sigma_{\text{Be-7}}}\right] \cdot \left[\frac{SF_{\text{H-3}}}{SF_{\text{Be-7}}}\right], \quad (1)$$

where $A_{\text{H-3}}$ and $A_{\text{Be-7}}$ are the measured specific activity of $^3$H and $^7$Be, respectively, at the end of the machine operation; $\sigma_{\text{H-3}}/\sigma_{\text{B-7}}$ is the ratio of cross sections of $^3$H and $^7$Be in the water target; and $SF_{\text{H-3}}$ and $SF_{\text{Be-7}}$ are saturation factors of $^3$H and $^7$Be, respectively, considering its decay and water-exchange history. Assuming that the cross sections of $^3$H and $^7$Be production in water are constant in the high-energy region, $\sigma_{\text{H-3}}/\sigma_{\text{B-7}} = 6.1$, which was found in the previous proton-irradiation study at 120 GeV [2], can be used to calculate the residual rate of $^7$Be.

The estimated residual rates of $^7$Be in the cooling water collected in December 2011 are listed in **Table 3**. The estimated residual rates of $^7$Be in (g) and (h) were approximately 5% for cooling water systems despite of the absence of a deionizer. It was assumed that approximately 95% of $^7$Be was either removed by the filter unit or it was attached to the surface of the pipes. The estimated residual rates in the other samples were less than 0.1%. The deionizers functioned to remove radionuclides from the cooling water as well as to maintain the desired electrical resistance of the cooling water. However, a small fraction of $^7$Be was still residual in the water. Weak retention of some radionuclides on the ion-exchange resin was observed in the cooling water system for the K2K magnetic horns [1]. Furthermore, colloid formation of radionuclides was observed in Cu-foil-soaked water bombarded with 120-GeV protons [2]. Therefore, the residual $^7$Be in the cooling water could have been due to weak retention of the $^7$Be colloids by the cation-exchange resin in the deionizer.

### 4. CONCLUSION

The observed variety of the radionuclides in the different cooling water systems indicated that behavior of radionuclides in cooling water complicatedly depends on both major and minor materials in contact with the water. Activity of radionuclides in cooling water depended on the presence of a deionizer. Specific activities of $^3$H could be used to estimate the residual rates of $^7$Be, which commonly had the highest specific activity among the γ-emitting radionuclides in the cooling water. The estimated residual rates of $^7$Be in the cooling water were approximately 5% for systems without deionizers and less than 0.1% for systems with deionizers. Although the deionizers function to remove radionuclides from the cooling water, $^7$Be activity was not completely removed by the deionizer. Our previous studies indicated that this could be due to the weak retention of $^7$Be colloids by the ion-exchange resins in the deionizer medium.

**Acknowledgements**

This work was supported by a grant-in-aid from the Ministry of Education, Science and Culture (KAKENHI 19360432 and 21360473) in Japan. Fermilab is a US Department of Energy Laboratory operated under Contract E-AC02-07CH11359 by the Fermi Research Alliance, LLC.

Table 3. Estimated residual rates of $^7$Be in the cooling water collected in December 2010.

| System name | Residual rate of $^7$Be |
|---|---|
| (a) NuMI, Target | 0.02 % |
| (b) NuMI, Magnetic horn1 | 0.02 % |
| (c) NuMI, Magnetic horn2 | 0.003 % |
| (d) NuMI, Decay pipe | 0.06 % |
| (e) NuMI, Hadron absorber | 0.06 % |
| (f) Pbar, Collection lens | 0.008 % |
| (g) Pbar, Pulse magnet and collimator | 4 % |
| (h) Pbar, Beam absorber | 5 % |